\documentstyle[12pt,aasms4]{article}

\lefthead{Colbert et al.}
\righthead{Large-Scale Outflows II}

\def\lapprox{\lesssim}
\def\gapprox{\gtrsim}
\def\etal{{et~al. }}
\def\asec{\arcsec}
\def\amin{\arcmin}
\def\han2{{H$\alpha$+[NII]}}
\def\deg{\arcdeg}

\begin{document}

\title{Large-Scale Outflows in Edge-on Seyfert Galaxies. \\
       II. Kiloparsec-Scale Radio Continuum Emission}

\author{Edward J. M. Colbert\altaffilmark{1,2},
        Stefi A. Baum\altaffilmark{1},
        Jack F. Gallimore\altaffilmark{1,2},
        Christopher P. O'Dea\altaffilmark{1},
        Jennifer A. Christensen\altaffilmark{1}
        }

\altaffiltext{1}{ Space Telescope Science Institute, 3700 San Martin Drive,
                  Baltimore, MD  21218  }
\altaffiltext{2}{ Department of Astronomy, University of Maryland,
                  College Park, MD  20742}

\begin{abstract}
We present deep images of the kpc-scale radio continuum emission in
14 edge-on galaxies (ten Seyfert and four starburst galaxies).
Observations were taken with the VLA at 4.9~GHz (6~cm).  The Seyfert galaxies
were selected from a distance-limited sample of 22 objects 
(defined in paper~I).
The starburst galaxies were selected to be well-matched to the Seyferts
in radio power, recessional velocity and inclination angle.
All four starburst galaxies have a very bright disk component and one
(NGC~3044) has a radio halo that extends several kpc out 
of the galaxy plane.  
Six of the ten Seyferts observed have large-scale (radial extent $\gapprox$1 
kpc) radio structures extending outward from the nuclear region, indicating
that large-scale outflows are quite common in Seyferts.
Large-scale radio sources in Seyferts are 
similar in radio power and radial extent to 
radio halos in edge-on starburst galaxies, but their morphologies do
not resemble spherical halos observed in starburst galaxies.
The sources have diffuse morphologies, but, in general, they
are oriented at skewed angles with respect to the galaxy minor axes.
This result is most easily understood if
the outflows are AGN-driven jets that are somehow diverted away from the
galaxy disk on scales $\gapprox$1 kpc.  Starburst-driven winds, however, 
cannot be ruled out.  More observational work is needed to
determine whether massive star formation is present at high enough rates
to drive galactic winds out to kpc scales in Seyfert galaxies.
\end{abstract}


\section{Introduction}

Optical light from quasars is primarily emitted by the active galactic
nucleus (AGN) and not by the host galaxy.  Similarly, radio emission from
classical radio galaxies is predominantly emission produced by the AGN.
However, in Seyfert galaxies, although the AGN
is often very luminous when compared with the host galaxy, it may not
always dominate the luminosity in a given waveband.  
For example,
in the archetypical Seyfert galaxy NGC~1068, a circumnuclear starburst 
is quite luminous at infrared (IR; Balick \& Heckman 1985) and X-ray 
(Wilson \etal 1992) wavelengths.  Rodriguez-Espinosa, Rudy \& Jones (1987)
concluded that most of the far-IR emission from Seyfert galaxies is due to
starburst episodes occurring in the galactic envelopes surrounding the
active nucleus.  This result, coupled with the known correlation between
far-IR and radio luminosity for starbursting systems 
(e.g. Helou, Soifer \& Rowan-Robinson 1985), suggests that much of
the radio emission
in Seyferts might also be produced by circumnuclear starbursts.

Early radio surveys 
(de~Bruyn \& Wilson 1976; Meurs \& Wilson 1981; Wilson \& Meurs 1982)
of Seyferts with the Westerbork Synthesis Radio Telescope 
demonstrated that, in most Seyferts, most of the radio flux is confined to 
the central region of the galaxy.  
Large radio structures analogous to jets in radio galaxies were not generally
detected and most of the detected objects were unresolved at this
resolution ($\sim$20\asec).  Soon after the 
Very Large Array (VLA\footnotemark[1])
\footnotetext[1]{
  The VLA is part of the National Radio Astronomy Observatory (NRAO), 
  which is operated by Associated Universities, Inc., 
  under contract with the National Science Foundation.
}
became available, Condon \etal (1982) imaged a sample
of bright spiral galaxies (some of which were Seyferts) with strong radio 
sources and argued that the radio emission from the central sources could
be explained by synchrotron radiation from supernova remnants (SNRs).  
However, Wilson \& Willis (1980) claimed that the AGN was producing much of
the nuclear radio emission.

Sub-kpc scale 
VLA surveys of Seyferts 
(see, e.g.,  Ulvestad \& Wilson 1989 and references therein) later showed that
nuclear sources in Seyferts come in several types of morphological classes.
`Linear' (L-type) radio structures $\lapprox$1 kpc are found in 
$\sim$20$-$30\% of Seyfert galaxies (Ulvestad \& Wilson 1989, Kukula 1993). 
`Diffuse' (D-type) central radio structures were also found,
at about the same frequency of occurrence.  Some Seyferts have 
have both L and D components.  Compact (unresolved, $\lapprox$1\asec)
nuclear radio sources 
were present in many of the objects.  Wilson (1988) suggested that the 
diffuse radio
components were produced by star-formation activity, while the compact and 
linear radio components were produced by the AGN.

The radio powers of Seyfert galaxies are
$\sim$10$^{22}$$-$10$^{25}$ W~Hz$^{-1}$ at centimeter wavelengths
(e.g.  Meurs \& Wilson 1984).  
Normal spirals (i.e., spirals which do not have AGNs) have radio powers 
which are, on average, several orders of magnitude lower
($\sim$10$^{20}$$-$10$^{23}$ W~Hz$^{-1}$; e.g., Hummel 1981).  
Hummel (1981) found that, in normal spirals, most ( $\gapprox$ 90\%) of
the total radio power comes from the disk component.  However, 
central radio sources were also found in some normal spirals.
Seyfert nuclei are usually
found in spiral hosts, so one would expect the total
radio emission from Seyfert galaxies 
to consist of at least (i) a central `nuclear' component produced by the AGN,
(ii) another central component, probably 
produced by SNRs in the nuclear region of the host galaxy (cf. Condon 1992),
and (iii), a `disk' component from the host galaxy.

A fourth component of radio emission in Seyferts has also been identified in
several objects.  Hummel, van Gorkom \& Kotanyi (1983) showed that, in 
large-scale\footnotemark[2]\
\footnotetext[2]{
  In this paper, the term `large-scale' refers to scales $\gapprox$1 kpc,
  unless otherwise noted.
}
maps of five edge-on galaxies 
(two were later identified as Seyferts and the other three as LINERs),
`anomalous' components extend out of the disk and appear to 
be coming from the nuclear region.
In a more recent study of 13 Seyferts selected for large-scale radio or optical
emission, Baum \etal (1993) found large-scale, 
extranuclear components in 12 objects and
showed that the radio sources were preferentially aligned
with the minor axes of the galactic disks.
They suggested that the sources could be produced by galactic winds powered
by circumnuclear starbursts and also discussed the possibility that they are 
produced by an AGN-driven outflow.  Although several cases of such
extraplanar emission have been noted in the literature, a systematic study
of these features in Seyfert galaxies has never been published.  One would
like to conduct such a study in order to
determine whether these large-scale radio structures are common in Seyfert
galaxies, understand their radio properties, and investigate whether they might
be produced by the AGN or circumnuclear star-formation activity.

This is paper~II from a study of large-scale outflows (LSOs) in a complete, 
distance-limited sample (see Colbert \etal 1996, hereafter paper~I)
of edge-on Seyfert galaxies.  In paper~I, 
we used \han2 images and/or minor-axis optical spectra of
22 edge-on Seyfert galaxies to show that LSOs
are likely to be present
in $\gapprox{{1}\over{4}}$ of all Seyferts.  In this paper, we present 
large-scale VLA maps of ten of the 22 galaxies in our complete sample.
Our maps allow us to search for emission at low surface-brightness levels 
($\gapprox$0.05 K at 6~cm), which is several times better than those of
most published large-scale radio maps of Seyferts.
We also present large-scale radio maps for a comparison sample of four 
edge-on starburst galaxies, selected to be well matched to the sample 
Seyferts.

Observations and data reduction techniques are described in section 2.
In section 3, we present results for individual objects and list
properties of the observed radio sources.  In section 4, we describe the
large-scale radio structures from Seyferts in our complete sample
and compare their properties with those of radio jets and lobes
in classical radio galaxies and radio halos in starburst galaxies.  
We then discuss AGN and starburst origins for the large-scale
radio structures.
A Hubble constant of 75 km~s$^{-1}$~Mpc$^{-1}$ is assumed throughout this paper.

\section{Observations and Data Reduction}

A total of 16 objects were observed (12 Seyferts and four starbursts).  
The Seyfert galaxy Mrk 577 was not detected.  We observed the Seyfert 
galaxy
IRAS~13197-1627 but later discovered it was not associated with the
edge-on spiral galaxy MCG~$-$3-34-63 (see Appendix for details and
radio maps of this object).  Thus, in this paper we present radio
maps for ten
edge-on Seyfert galaxies and four edge-on starburst galaxies.

Observations were made at 6~cm (central frequency 4860.1 MHz), with the 
VLA on 1994 June 12, while the array was in C-configuration.  
The effective bandpass was 100 MHz and integration times were $\sim$45 min 
per galaxy.  

Calibration and reduction of the data was performed using the NRAO AIPS data 
reduction package.  Visibility data were flux density calibrated
from observations of 3C286 and phase calibrated using frequent observations 
of nearby VLA point-source calibrators.  

For all objects detected, the visibility data
were improved using several cycles of self-calibration and mapping
(cf. Perley \etal 1989).
The shortest spacings are $\sim$1 k$\lambda$, 
so the images are sensitive to structure $\lapprox$3\amin\ in size.

Several types of CLEANed maps were made from the final self-calibrated 
visibilities.  All maps presented in this paper have been 
restored with circular
Gaussian beams.  Circular beams were chosen over elliptical beams so that
we could measure geometrical properties of the extended emission directly from
the maps.  Maps with beam size 5\asec\ FWHM (full-width half-maximum),
the approximate size of the synthesized beam, have sensitivities 
$\sim$20$-$25 $\mu$Jy/beam, which 
corresponds to a brightness temperature of $\sim$50 mK.  These maps are
hereafter referred to as the ``normal'' maps.  Next, a set of ``hi-res'' 
maps were made by restoring the image with a circular Gaussian
beam with FWHM $=$ 3\asec.  These maps were used for determining the 
structure of the nuclear emission and for estimating the nuclear radio flux.
Finally, we used a 20~k$\lambda$ Gaussian taper on the visibility data to
produce ``tapered'' maps (circular beam, FWHM $=$ 10\asec), 
which are more sensitive to large-scale structure.
The normal and tapered maps were made using natural weighting and the hi-res
maps were made using uniform weighting.

Archival data from observations described in Edelson (1987) were retrieved 
from the VLA archives, recalibrated and reduced to maps using self-calibration
and CLEANing techniques.
These data were taken at both 1.46 and 4.89 GHz (20 and 6 cm),
when the array was in D-configuration. 
Although the beam sizes in these maps are considerably larger 
($\sim$20\asec\ and 60\asec\ FWHM, respectively), the maps are sensitive 
(rms noise at 6~cm $\sim$ 0.1 mJy~beam$^{-1}$ $\sim$ 20 mK)
to 
structure $\gapprox$3\amin\ and so were used to supplement the 6~cm 
C-configuration maps.  All of the maps of the Seyferts in our complete sample
(paper~I) are unresolved in these data, except for those of NGC~4235.  
A low-resolution 1.46~GHz map of 
NGC~4235 from this D-configuration data is presented in section 3.

\section{Results}

A list of the 22 edge-on Seyferts in our complete distance-limited sample 
is given in Table 1.  The ten Seyferts for which we have obtained new radio 
maps are marked with a dagger (\dag).  
Host galaxy types, optical positions, axial ratios and 
recessional velocities for these galaxies can be found in Table 1 of paper~I.  
Here we list assumed distances to the galaxy, total 6~cm radio power P$_6$,
total far-infrared (FIR) luminosity L$_{FIR}$ and $\mu$, the logarithm of
the ratio of 60$\mu$ flux to total 6~cm radio flux.

The mean value of $\log$ P$_6$ for 
our complete sample of Seyferts is 21.4 $\log$~W~Hz$^{-1}$, which is
comparable to total radio
powers for larger samples of Seyferts (e.g. Edelson 1987, Kukula \etal 1995). 
The Seyferts observed by Baum \etal (1993) are, on average, higher 
in total radio power
by a factor of $\sim$3 (mean value of $\log$ P$_6 =$ 21.8 $\log$~W~Hz$^{-1}$),
which can be explained by their selection criterion to observe Seyferts that
previously had evidence for large-scale radio emission.  Since our sample of
Seyferts was selected by distance, our results are more representative for
all Seyferts, not just radio-loud ones.

In Table, 2, for each of the four edge-on starburst galaxies,
we list galaxy morphologies,
axial ratios, recessional velocities, assumed distances, P$_6$, L$_{FIR}$ 
and $\mu$.
These galaxies were selected to be well-matched to the Seyferts in radio 
luminosity, axial ratio and recessional velocity. 

Literature searches were done for all 26 galaxies (22 Seyferts and 
four starbursts).  We did not find any
information about possible radio emission from the Seyfert 
galaxies ESO~103-G35, NGC~6810 and IC~1417.  Radio observations of
the other galaxies are discussed in the following two subsections.
In section 3.1, we present our new radio maps of ten of the 22 Seyfert 
galaxies and discuss results from previous radio observations.
In section 3.2, we present new radio maps 
(and discuss previous radio observations)
of the four starburst galaxies.

\subsection{Individual Objects: Seyfert Galaxies}

\subsubsection{IC~1657}

We did not obtain radio images of IC~1657.  Unger \etal (1989) observed
this galaxy with the VLA in a hybrid B-C configuration, 
but do not show a map of the radio emission.  They quote a total flux of 21 mJy 
at 1.4~GHz. 

\subsubsection{UM~319}

In Figure 1a, we show contour maps 
of the radio emission from UM~319 overlaid on an optical image.
The large-scale structure extends out to $\sim$10\asec\ (3 kpc) 
to the southeast and out to $\sim$5\asec\ to the northwest, which is not
quite out of the disk, in projection.  From the hi-res map, one can notice
that the radio morphology roughly appears linear in structure, extending along
P.A. $\approx$ 125\deg.  
Such a `linear' radio morphology is suggestive of a bi-symmetric nuclear
outflow, but follow-up observations at higher resolution are necessary 
to determine the radio morphology on smaller scales.  Additional observations 
at lower-surface brightnesses (or lower frequencies, if the extranuclear 
component has a steep spectrum) would
be useful for determining if the observed extranuclear emission extends
out further from the nucleus.

\subsubsection{Mrk~993}

A contour map of the radio emission from Mrk~993 is shown over an optical 
image of the galaxy in Figure 1b.  The central radio source is unresolved 
in our maps.  It 
is also unresolved in 8.4~GHz
maps (VLA, A and C configuration) presented in
Kukula \etal (1995).

\subsubsection{Mrk~577}

We observed Mrk~577 but did not detect it (3$\sigma$ upper limit 
0.12 mJy at 4.9~GHz).
Previous workers (e.g., Kojoian \etal 1980, Mulchaey \etal 1995) have observed 
this galaxy, but also failed to detect it.

\subsubsection{Ark~79 (UGC~1757)}

Contour maps of the radio emission from Ark~79 are shown in
Figure 1c.  The nuclear radio source is slightly
resolved (see hi-res map) and extends southward and slightly
westward along the disk.  The extranuclear emission does not quite extend out
of the plane of the disk (in projection), but does extend in the general
direction of the minor axis.  No small-scale maps of the 
nuclear radio emission from Ark~79 have been published.

\subsubsection{NGC~931 (Mrk~1040)}

We did not obtain radio images of NGC~931. Rush, Malkan \& Edelson (1995)
find a flux of 6.2 mJy at 4.9~GHz (VLA, D-configuration), but do not show
a map.

\subsubsection{NGC~1320 (Mrk~607)}

Contour maps of the radio emission from NGC~1320 are shown in Figure 1d.
In addition to the nuclear radio source, we find emission extending
out to $\sim$5\asec\ (0.9 kpc) to the south from the
nucleus in P.A. $\approx$180\deg\ (see hi-res map).  
The emission also extends slightly northward in the same P.A., and also
to the northwest, along the disk.
As in UM~319, the emission appears linear but does not extend out of the
disk (in projection).  Follow-up radio observations would be useful for
determining if structure is present on smaller and larger scales.

\subsubsection{NGC 1386}

We did not observe NGC~1386.  No extranuclear radio emission
is resolved in the large-scale 1.4~GHz radio map presented in Condon (1987).
Rush \etal (1995) quote a flux of 21.5 mJy at 4.9~GHz from VLA D-array 
observations.  On small scales, the nuclear radio source is only slightly
extended (out to $\sim$400 pc) along the galaxy major axis 
(southwest along P.A. 55\deg, 
Weaver, Wilson \& Baldwin 1991; Ulvestad \& Wilson 1984).
Weaver \etal (1991) present evidence for a 
nuclear outflow directed along the {\it major} axis.

\subsubsection{NGC 2992}

In Figure 1e, we show contour maps of the radio emission from NGC~2992.
The nuclear radio source is
surrounded by fainter emission with complex structure extending in nearly all
P.A.'s on a scale of $\sim$20$-$30\asec\ (3$-$4.5 kpc).  Extraplanar emission 
to the east and west is noticeable in the normal and tapered images. 
The large-scale 20~cm map from Hummel \etal (1983, reproduced in Figure 1e;
see also map in Ward \etal 1980) clearly shows extraplanar emission 
extending to the east, but this emission is not well imaged in 
our 4.9~GHz (6~cm) maps.  By comparing Hummel et al.'s map
with a map made from our data which has the same beam size, we find that
the eastern `arm' of radio emission has a very steep spectrum 
($\alpha \gapprox$ 2; S$_\nu \propto \nu^{-\alpha}$).
On smaller scales, the nuclear morphology consists of a compact source plus
surrounding diffuse emission (Ulvestad \& Wilson 1984) which resembles a
``striking pair of loops'' extending $\sim$2\asec\ (0.3 kpc) north 
(and south) from the compact nuclear source in P.A. $\sim$0\deg
(Wehrle \& Morris 1988).  

\subsubsection{MCG~$-$2-27-9 }

We show a contour map of the radio emission from MCG~$-$2-27-9 (overlaid on 
an optical image) in Figure 1f.  The central radio source is unresolved.
However, we detect emission from a source which is positioned $\sim$35\asec\
(10.5 kpc at the distance of MCG~$-$2-27-9)
to the south from the nuclear source, along the galaxy minor axis.  This source
has flux density 0.3 mJy, which corresponds to
P$_6 =$ 1.3 $\times$ 10$^{20}$ W~Hz$^{-1}$ at the distance
of MCG~$-$2-27-9.  Follow-up observations would be useful to determine if this
source is associated with this galaxy.

\subsubsection{NGC~4235}

In Figure 1g, we show contour maps of the radio emission from NGC~4235.
Diffuse structures extend out $\sim$1\amin\ 
(9.3 kpc) from the nucleus, both eastward and westward from the nuclear region.
The morphology of the 
extraplanar emission appears diffuse and bubble-like, especially toward
the west.  At large radii, the eastern source may be slightly more extended 
toward the galaxy minor axis (see low-resolution 20~cm map).

The nuclear radio source
is unresolved in small-scale VLA maps 
(Ulvestad \& Wilson 1984; Kukula \etal 1995).

\subsubsection{NGC~4388}

We did not obtain new radio images of NGC~4388.  Large-scale radio
maps (Condon 1987; Hummel \etal 1983) reveal diffuse emission 
extending out along the minor axis from the nuclear region.  In Figure 1h,
we reproduce a contour map (from Hummel \etal 1983) of the 5~GHz radio 
emission which shows the large-scale radio structures.  In a
higher-resolution radio image by Stone, Wilson \& Ward (1988), one can resolve
individual `fingers' extending out of the disk into the halo.

\subsubsection{NGC~4602}

We show a contour map of the radio emission from NGC~4602 
in Figure 1i.  The nuclear source is quite weak and most of the
emission comes from sources in the galaxy disk.

\subsubsection{NGC~4945}

The large-scale 4.75~GHz map from Harnett \etal (1989) is reproduced in
Figure 1j.  The extraplanar structure in NGC~4945 extends $\sim$10\amin\ 
(20 kpc) along the minor axis on both sides of the disk.

\subsubsection{IC~4329A}

We did not obtain new radio images of IC~4329A.  As far as we know, the
only published radio maps of this galaxy are those in Unger \etal 
(1987, see Figure 1k).  Diffuse radio structure extends westward 
$\sim$6\asec\ (1.9 kpc) in P.A. $\sim$285\deg, in the general direction of
the galaxy minor axis.

\subsubsection{NGC~5506}

We show contour maps of the large-scale radio emission from NGC~5506
in Figure 1l.  Diffuse, bubble-like radio structures extend out of the disk
to radii $\sim$30\asec\ (3.6 kpc) from the nucleus in 
P.A. $\sim$140\deg.  At higher resolution, the nuclear source resolves into a
compact core plus diffuse halo 
(Ulvestad, Wilson \& Sramek 1981; Unger \etal 1986, 1987) extending
in the same direction as the large-scale `bubbles.'   
Wehrle \& Morris (1987) have mapped the small-scale radio halo and have 
identified a `loop' of diameter $\sim$100~pc which originates at 
the compact source.  They attribute this loop to either
a bubble of hot plasma rising
from the nucleus or a magnetically dominated coronal arch.  

\subsubsection{IC~1368}

A contour map of the radio emission is shown in Figure 1m.
The nuclear source is very slightly resolved in P.A. $\sim$ 45\deg, in
the direction of the major axis.

\subsubsection{NGC~7410}

Condon (1987) has
published a large-scale 1.5~GHz radio map, which shows a bright, extended
radio structure positioned roughly 2.5\amin\ (17.0 kpc) to the northwest,
along the minor axis (see Figure 1n).  The extended source has
a 1.5~GHz flux of 6.3 mJy, which corresponds to a 6~cm radio power
P$_6 =$ 1.7 $\times$ 10$^{20}$ W~Hz$^{-1}$ (assuming spectral index 
$\alpha =$ 0.75).  Its connection with NGC~7410 is speculative, 
but follow-up studies are warranted to determine if it
is associated with this galaxy.

\subsubsection{NGC 7590}

We did not obtain radio images of NGC~7590 
and we are not aware of any published radio 
maps of this galaxy.  Wright (1974) quote a flux density of 70 mJy at 1.4~GHz
(and 2.7~GHz) from single-dish observations and Ward \etal (1980) show that 
the radio spectrum is quite flat.

\subsubsection{Possible Supernova Remnants or Radio Supernovae in the Disks of
Seyfert Galaxies}

Isolated, unresolved radio sources were detected in the disks of the 
Seyfert galaxies UM~319, Ark~79, NGC~1320 and MCG~$-$2-27-9.  
Such radio sources may also be present in the disk of the Seyfert
galaxy NGC~4602 (see Figure 1i); however, it is difficult
to determine from our map if they are individual sources 
or just localized peaks of diffuse emission from the disk.  In Table 3,
we list positions, 6~cm fluxes and 6~cm radio powers for the disk sources
in these five galaxies.

Typical radio powers are $\sim$1$-$8 $\times$ 10$^{19}$ W~Hz$^{-1}$,
which is at least a factor of 10 larger than the radio power of Cas~A
(P$_6$[Cas~A] $=$ 0.8 $\times$ 10$^{18}$ W~Hz$^{-1}$).  However, the
radio powers are comparable to those of radio
supernovae (see, e.g. Sramek \& Weiler 1990).  
Follow-up radio studies of all of these galaxies
would be useful for determining if the unresolved sources are individual radio 
supernovae, or perhaps groups of SNRs.

\subsection{Individual Objects: Starburst Galaxies }

\subsubsection{UGC 903 (MCG~3-4-26)}

In Figure 2a, we show a contour map of the radio emission from UGC~903.
Much of the radio emission comes from a disk component.
Faint `fingers' appear to extend out of the disk, but it is not clear whether
this emission originates 
from regions above the galaxy plane or is beam-smeared emission
from point-sources in the disk.

\subsubsection{NGC 1134 (Arp~200)}  

A contour map of the radio emission is shown in 
Figure 2b.  Nearly all of the emission comes from a 
box-like region surrounding the nucleus where massive stars are apparently
forming.  Such boxy morphology is also present in the 1.5~GHz map from 
Condon \etal (1990) and the \han2 image from Lehnert \& Heckman (1995).

\subsubsection{NGC 3044}

In Figure 2c, 
we display contour maps of the radio emission from this galaxy.
As in the other starburst galaxies in our sample, much
of the radio emission from this galaxy comes from the disk component.
The disk in NGC~3044 is highly inclined and it is easy to confirm the presence 
of a radio halo extending above the disk
(e.g., see our tapered image).  This radio halo has been previously
noted by Hummel \& van der Hulst (1989). 
The 1.5~GHz map from 
Condon \etal (1990, reproduced in Figure 2c) clearly shows
emission from this radio halo.

\subsubsection{NGC 7541}

We show a contour map of the radio emission overlaid on an 
optical image in Figure 2d.  Again, most of the emission comes from 
sources in the disk, especially from regions along the eastern arm.
As in UGC~903, `fingers' seem to extend out of the disk, but this may be
beam-smeared emission from point sources in the disk.

\subsection{Properties of the Large-Scale Radio Sources}

\subsubsection{Radio Fluxes and Powers}

For each of the 14 galaxies we observed, we list the total 4.9~GHz radio flux
(F$_T$) in Table 4, column 2 (corresponding total radio powers are 
listed in Table 1).  Total fluxes were measured (using the normal maps)
by summing the flux density inside boxes enclosing all of the emission from
the galaxy. We also list nuclear fluxes F$_N$ 
(peak flux density in the nuclear region in the hi-res maps)
and corresponding nuclear powers P$_N$ (Table 4, columns 3 and 4, respectively).

One would like to separate the radio emission produced by the outflow
from that produced by the rest of the galaxy.  In order to achieve this,
we first calculated an
`extranuclear' radio power P$_{XN}$ (Table 4, column 5) 
from the extranuclear flux F$_{XN} =$ F$_T -$ F$_N$.
We then corrected P$_{XN}$ by subtracting 
radio powers of sources that were clearly associated with disk emission
(see, e.g., Table 3).  In most cases, the total radio power from individual
sources in the disk is much smaller than that from the LSO.
These `corrected' extranuclear powers 
are good estimates of the radio power from the LSOs, so
we denote them as P$_{LSO}$ in Table 4 (column 6).

This method worked quite well for most of the Seyfert galaxies since their
disk sources are, in general, isolated from the nuclear radio sources.
In three Seyferts (Mrk 993, MCG~$-$2-27-9 and IC~1368) the extranuclear
emission is not well imaged in our maps and we do not list values for 
P$_{LSO}$.  In the four starburst galaxies and in the Seyfert galaxy NGC~4602,
the disk emission is not easy to distinguish from possible extraplanar 
emission.
Although extraplanar radio emission is clearly present in the starburst
galaxy NGC~3044, estimating the radio power of this
component is not straightforward.  An accurate estimate would involve 
successfully modeling the radio morphology of the disk component.  
We estimate a rough extraplanar flux of 0.3 mJy (from Figure 2c),
which corresponds to a 4.9~GHz radio power of 10$^{19.7}$ W~Hz$^{-1}$
($\sim$10$^{-1.4}$ of the total 6~cm radio power).
Thus, we were able to measure P$_{LSO}$
for six of our Seyfert galaxies and one of our starburst galaxies.

\subsubsection{Radial Extent from Nucleus and Orientation}

Geometrical properties were measured for the large-scale radio sources in 
nine of the 22 Seyfert galaxies in our complete sample.  Other than the
six Seyferts discussed in the previous section, three more galaxies in our
complete sample (NGC~4388, NGC~4945 and IC~4329A, which we did not observe)
have large-scale radio structures.
Although NGC~7410 and MCG~$-$2-27-9 both have radio sources which are 
positioned along their minor axes, 
more evidence is needed to show that the sources are associated with
these galaxies.
In Table 4, for each of these nine galaxies, we list
position angle (P.A.) with respect to the nucleus,
maximum projected radius from the nucleus, P.A. of the
major axis of the galactic disk, and $\Delta$, the difference in angle between 
the P.A. of the radio source and that of the major axis.

\subsubsection{Intrinsic Properties}

For the six Seyfert galaxies in Table 4 with large-scale radio structures,
we estimated maximum lifetimes (cf. van der Laan \& Perola 1969) of
the cosmic rays producing the synchrotron radiation.
Equipartition between magnetic field and particle energy was assumed and 
the radio sources were assumed to have cylindrical geometry.  Magnetic 
pressures in the radio sources are $\sim$0.5$-$10 $\times$ 10$^{-14}$ 
erg~cm$^{-3}$ and total energies (in particles and magnetic field) are 
$\sim$0.1$-$2 $\times$ 10$^{53}$ erg.
By dividing the lifetimes (typically 1$-$2 $\times$ 10$^7$ yr) of the
synchrotron-emitting electrons by the radial extent of the sources (Table 5),
we estimate {\it minimum} outflow velocities of $\sim$50$-$100 km~s$^{-1}$ 
for UM~319, Ark~79, NGC~1320, NGC~2992 and NGC~5506 and $\sim$450 km~s$^{-1}$
for NGC~4235.  Thus, near-relativistic outflows are not required.

\section{Discussion}

\subsection{The Nature of Large-Scale Radio Emission in Seyfert Galaxies}

\subsubsection{Frequency of Occurrence}

Large-scale radio structures seem to be a fairly common feature of Seyfert
galaxies.
In six of the ten Seyferts we observed, large-scale radio sources clearly
extend out of the nuclear region (i.e. radii $\gapprox$1 kpc).  Therefore,
such large-scale radio structures are 
likely to be present in $\gapprox$${{1}\over{2}}$ of all Seyferts.
Suitable radio maps are not available for many of the remaining 12 objects in 
our complete sample, but such large-scale radio emission is clearly present in 
at least three of these 12 galaxies.

In comparison, in starburst galaxies, large-scale radio emission 
at these surface brightness levels is probably less common.
Although our detection frequency of ${{1}\over{4}}$ is consistent with an
inherent frequency of occurrence of 50\%, in larger surveys of edge-on
galaxies, only a few percent of edge-on 
galaxies show signs of radio halos (e.g. Hummel, Beck, \& Dettmar 1991).

\subsubsection{Radio Powers}

The 6~cm radio powers P$_{LSO}$ of the large-scale radio sources in
the Seyferts are
$\sim$10$^{20}$$-$10$^{22}$ W~Hz$^{-1}$ (Table 4), which are at least
several orders of
magnitude weaker than radio powers of lobes in typical FR~II radio galaxies.
It is also noteworthy that, while in radio galaxies core powers are weaker
than lobe powers (see, e.g., Zirbel \& Baum 1995), in Seyferts the nuclear
powers P$_N$ are typically larger than P$_{LSO}$ 
(Table 4; see also Baum \etal 1993).

Measurements of radio powers of halos in starburst galaxies are not widely
available, due to the difficulty of separating emission from the disk.
Our estimate of the radio power of the halo in NGC~3044
is $\sim$10$^{19.7}$ W~Hz$^{-1}$ (section 3.3), roughly consistent with the
range of P$_{LSO}$ for our Seyferts.

\subsubsection{Radial Extent from Nucleus}

In Figure 3, we show a histogram of the maximum projected radius of
the large-scale radio emission in the Seyferts.  Typical radii
are 1$-$5 kpc, although large-scale structures extend out to $\sim$10 and
$\sim$30 kpc in NGC~4235 and NGC~4945, respectively.
Typical sizes for lobes in radio galaxies are larger, up to
$\sim$10$^2$$-$10$^3$ kpc.  
Radial extents of radio halos in starburst galaxies are more typical of what is
found for large-scale sources in Seyferts.
For example, the radial
extent of the radio halo in NGC~3044 is $\sim$2.5 kpc (Figure 2c).
More prominent halos, such as those in NGC~253 (Carilli \etal 1992), M82
(Seaquist \& Odegard 1991) and NGC~4631 (cf.  Dahlem, Lisenfeld \& Golla 1995),
extend out to radii of several tens of kpc.

\subsubsection{Morphology}

Although large-scale 
radio sources in Seyferts are similar in radio power and 
radial extent to those in starburst galaxies, their radio morphologies do not
typically resemble spherical halos.  An exception is the extended radio source
in the Seyfert galaxy NGC~4945 (Figure 1j).  According to 
Heckman, Armus \& Miley (1990),
a starburst-driven superwind is present in this galaxy and so, in this case, 
the radio halo may be produced by this superwind.
In general, large-scale radio sources in Seyferts are diffuse
structures originating from a small region surrounding the nucleus.
In some objects, such as UM~319 and NGC~1320, the structures
have linear morphologies which resemble collimated outflows. 
Others, such as NGC~4235, have bubble-like radio structures.

\subsubsection{Orientation}

At what angle do the radio sources emerge from the galaxy disk?
In Figure 4, we plot a histogram of $\Delta$ (Table 5, column 6), the
difference
between the P.A. associated with the extended radio
emission (Table 5, column 2) and the P.A. of the galaxy major axis (Table 5,
column 5).  
Radio halos in edge-on starburst galaxies are typically oriented 
perpendicular to the galaxy major axes.
Therefore, if LSOs in Seyfert galaxies are wide-angled winds blowing 
out perpendicular to the galaxy disk,
one might expect the distribution in $\Delta$ to be peaked near 90\deg.
If, on the other hand, the LSOs are jets that are
oriented isotropically, the distribution in $\Delta$ is expected to be flat.
The observed distribution in $\Delta$ (Figure 4) is consistent with directed
outflows (e.g., jets) that,
for some reason, do not extend to kpc scales
when they are oriented with the plane of the galaxy disk ($\Delta \sim$ 0\deg),
but are otherwise oriented isotropically.

Baum \etal (1993) found that the extranuclear sources in their 
sample were preferentially oriented perpendicular to the major axes
($\Delta \gapprox$ 60\deg).  Many of the galaxies in Baum et al.'s sample
were relatively face-on, which can give rise to misleading results for
the following two reasons.
First, the P.A. of the major axis is difficult to determine and thus has larger
uncertainty.  Secondly, in face-on galaxies, the relationship 
between $\Delta$ 
and the angle between the radio source and the plane of the galaxy disk is
more ambiguous.  Thus, the results from our sample of
edge-on Seyferts may be more accurate.
However, another consideration is that, as previously noted, 
the galaxies in Baum et al.'s sample have larger total radio power than those 
in our sample.  
More powerful outflows will more easily penetrate the galaxy disk and extend
to larger radii.  Thus, they may be
more easily influenced by global forces from
the galaxy (i.e. buoyancy forces) than are weaker outflows which do not 
escape from the nuclear region.

\subsubsection{Associated Optical Emission}

In paper~I, we presented \han2 images and optical long-slit spectra of 22 edge-on
Seyfert galaxies and used these data and other published data to search for
evidence for LSOs.  Our results implied that LSOs are present
in $\gapprox$ ${{1}\over{4}}$ of all Seyfert galaxies.  From our new radio
data, we find that that large-scale radio emission in Seyfert galaxies is 
even more common (present in $\gapprox{{1}\over{2}}$ of objects observed).

In Table 6, we list all objects in our complete sample of Seyfert galaxies 
which have evidence for an outflow from either optical or radio observations.
Ten of 22 galaxies 
are listed, consistent with LSOs being present in
$\gapprox$ ${{1}\over{2}}$ of all Seyferts.
Extraplanar radio emission and minor-axis optical emission-line regions
are fairly common, but obviously the presence of one does not always imply
the presence of the other.  Optical line emission produced by shocks in 
superwinds typically has low surface-brightness 
(10$^{-17}$ erg~s$^{-1}$~cm$^{-2}$~arcsec$^{-2}$ or lower, see, e.g.,
 Heckman, Armus \& Miley 1990) and could easily have been present, but 
below our detection limits.  On the
other hand, some of the optical line emission could be photoionized by 
the AGN, in which case it may not be co-spatial with radio emission produced
by the outflow, if such radio emission is present.

Of the nine galaxies with 
large-scale radio sources, five 
(NGC~2992, NGC~4388, NGC~4945, IC~4329A and NGC~5506)
also
show strong optical evidence for an outflow.  These four galaxies also show
kinematically disturbed (double-peaked and complex line profiles) and
high velocity gas along their minor axes (Tsvetanov, Dopita \& Allen 1995;
Corbin, Baldwin \& Wilson 1988; Heckman, Armus \& Miley 1990; 
Wilson \& Penston 1979 [nuclear spectra only]; 
Wilson, Baldwin \& Ulvestad 1985, respectively).

An interesting case in which we did not find associate optical line emission
is that of NGC~4235.  
Radio emission clearly extends along the minor axis, on both
sides of the galaxy plane (Figure 1g), but extended optical line emission 
is present along the {\it major} axis of the galaxy (paper~I; Pogge 1989).

Conversely, some objects show extended optical emission-line gas along their 
minor axis without associated radio emission.  For example,
in paper~I, we showed that in IC~1368, a bi-symmetric emission-line 
halo extends along the minor axis, but our radio maps do not show any 
extraplanar radio emission extending in that direction (see Figure 1m).

\subsubsection{Summary of Properties}

In summary, large-scale ($\gapprox$1 kpc) radio sources in Seyferts 
(1) are quite common (present $\gapprox$${{1}\over{2}}$ of the time);
(2) typically have 4.9~GHz
radio powers $\sim$10$^{20}$$-$10$^{22}$ W~Hz$^{-1}$, lower than that
of jets and lobes in radio galaxies, but approximately the same 
as that of radio halos in starburst galaxies; 
(3) typically extend out to radii of $\sim$1$-$5 kpc;
(4) have diffuse morphologies resembling emission from
collimated outflows from the nuclear region;
(5) appear to emerge from the disk at angles $\sim$40$-$90\deg\ with respect
to the galaxy major axes; and
(6) are often (but not always) associated with optical line emission and
kinematically disturbed gas.
 
\subsection{The Origin of Large-Scale Radio Emission in Seyfert Galaxies}

Since we have established that large-scale ($\gapprox$ 1 kpc) radio sources
are fairly common in Seyferts, it is important to determine
how they are produced.
A natural explanation 
is that the Seyfert nucleus both produces the relativistic plasma
and powers the outflow, perhaps in a `jet' analogous to those found in 
classical radio galaxies and radio-loud quasars.  The jet would then interact
with the galactic medium, shocking clouds of gas and entraining gas as it
moves outward, producing associated optical and X-ray emission.    
However, relativistic plasma can also be produced by star-formation
activity (e.g., SNRs) in the nuclear region, and, if a strong circumnuclear
starburst is present, it could also provide energy for driving an outflow 
out of the nuclear region.  Thus, two central questions to ask are
(1) What produces the relativistic plasma? and (2) What energy source
powers the kpc-scale outflows?

\subsubsection{Jets from the Active Nucleus}

A small subset of Seyfert galaxies are known to have nuclear jets
(e.g. Ulvestad, Neff \& Wilson 1987 [NGC~1068], 
Harrison \etal 1986 [NGC~4151], and Kukula \etal 1993 [Mrk~3]).  One would
like to know if such jets are ubiquitous in Seyferts and if
they are associated with the large-scale radio structures.

If the large-scale radio structures are from jets, how
does one explain the fact that
large-scale radio structures in Seyferts are not, in general, aligned with 
nuclear `linear' radio structures (e.g. Baum \etal 1993)?  One possibility
is that sub-kpc jets are
diverted by dense gas clouds in the nuclear region.  Such
a scenario has been proposed for the bending of the radio jet in NGC~1068
(Gallimore, Baum \& O'Dea 1996).  If the
jet manages to plow through the dense nuclear environment and continues
to flow outward, it might then be affected by ram pressure from rotating
gas in the galaxy disk.  This effect may be responsible for
the Z-shaped optical and radio structures in NGC~3516
(Miyaji, Wilson \& P\'{e}rez-Fournon 1992; Mulchaey \etal 1992).  
Jets which are deflected by gas clouds or ram pressure from
the galaxy disk will not emerge from the nuclear region at $\Delta \approx$
0\deg.  This is consistent with our result that, for our sample Seyferts,
$\Delta \gapprox$ 40\deg\ (section 4.1.5).  If jets in
Seyferts are, in general, weaker than those in radio galaxies, they may
lose enough momentum so that they never make their way out of the 
nuclear region.  This may explain the relatively high ``core-to-jet'' ratios 
we found for our Seyfert galaxies (section 4.1.2).

On the other hand, if the outflows do have sufficient momentum to blow out
to larger radii (perhaps as a wind),
buoyancy forces may align the outflows so that they continue
flowing out roughly perpendicular to the galaxy disk. 
This may explain why the large-scale radio sources in Baum et al.'s Seyferts
(which had larger total radio powers than ours)
were apparently preferentially oriented along the galaxy minor axes.
A specific example of where buoyancy forces may have redirected the outflow 
can be found in NGC~4388.  The (sub-kpc) linear nuclear
radio source is oriented at P.A. $\approx$ 25\deg.  Saikia \& Hummel (1989) 
show that
diffuse emission extends from this linear source in the direction of the
minor axis (P.A. 0\deg).    This diffuse emission is most likely connected 
with the large-scale radio structures, which are oriented along the same 
P.A. (see Figure 1h; Stone, Wilson \& Ward 1988).

Another possibility is that the AGN drives a wide-angled wind.
If the jet flow is stopped in the nuclear region, the energy from the jet
will rapidly thermalize and a wind
with properties very similar to those of a starburst-driven `superwind'
could be driven outward.

\subsubsection{Outflow from a Nuclear Starburst}

There seems to be good evidence that nuclear star formation is fairly common 
in Seyferts 
(NGC~1068: Balick \& Heckman 1985; 
NGC~4388: Corbin, Baldwin \& Wilson 1988, Stone, Wilson \& Ward 1988;
NGC~4945: e.g. Heckman, Armus \& Miley 1990;
NGC~7469: Wilson \etal 1986, 1991;
Mrk~231: Hamilton \& Keel 1987, Lipari, Colina \& Macchetto 1993;
Rodriguez-Espinosa \etal 1987).  
Therefore, to some degree, relativistic plasma will be present in the nuclear
region, in the SNRs themselves and in the surrounding interstellar medium.  
If a strong nuclear starburst is present, it 
could drive a wide-angled wind out along the galaxy rotation axis, away from
the nuclear region.  Such a wind will carry some of the plasma (including any 
produced by the active nucleus) out to kpc scales above the disk.  
If the nuclear star
formation is not strong enough to power a wind, star formation could still
be an important influence on the large-scale radio sources, as the
relativistic plasma produced by the starburst could be carried out to kpc
scales by an AGN-driven jet.

Assuming the supernova rate scales with non-thermal radio power,
we can estimate upper limits for the supernova rates for putative starbursts
in Seyferts.  Radio powers of 
P$_6 \sim$ 10$^{20.3}$$-$10$^{22.3}$ W~Hz$^{-1}$
(Table 1) corresponds with supernova rates of 
$\sim$ 10$^{-2.3}$$-$10$^{-0.3}$ yr$^{-1}$ (cf. Condon \& Yin 1990).
Most of the radio
emission in Seyferts originates from the nuclear region and, in some cases,
D-type sources, which are probably associated with star 
formation (Wilson 1988), dominate the nuclear radio luminosities.  Thus,
it is feasible that starburst-driven winds may be present is some Seyferts.  
As an example, a starburst with a supernova rate of 
10$^{-2.3}$ yr$^{-1}$ will drive a
galactic wind with kinetic luminosity $\sim$10$^{41.2}$ erg~s$^{-1}$
(cf. Heckman, Armus \& Miley 1990), which is consistent with our estimates
of kinetic luminosities of LSOs from paper~I.

As mentioned previously, in general, 
the large-scale radio sources in the Seyferts in our sample do not have 
spherical halo-like morphologies, as in starburst galaxies.  Furthermore,
the radio sources are
not strongly oriented along the minor axes (i.e., $\Delta \sim$ 40$-$90\deg).
Hence, if
the outflows are starburst-driven winds, either they are collimated somehow as
they emerge from the nuclear region, or it is only that the radio structures
appear to be from collimated outflows.
An important distinction to make between winds in starburst galaxies and
those envisaged for Seyferts is that the winds in Seyferts would originate
from a much smaller region around the nucleus
($\lapprox$ 1 kpc, compared to several kpc in starburst galaxies).  Thus, at 
small radii, the radio morphology of a starburst-driven wind in a
Seyfert may resemble
that from a collimated outflow.  On slightly larger
scales, circumnuclear tori of dense gas clouds, which
are known to be present in some Seyferts (e.g., the kpc-scale molecular ring
in NGC~1068), may collimate a starburst-driven wind by blocking gas which
flows out along the disk.  If the circumnuclear gas clouds are not uniformly 
distributed in a torus, the outflow could emerge from the nuclear
region at a skewed angle with respect
to the galaxy disk.  For example, if a wind encounters individual gas clouds 
which are positioned such as to directly block the outflow perpendicular
to the galaxy disk, the radio 
structure of the wind might resemble what we observe in the Seyferts: diffuse 
morphologies which are not preferentially oriented along the galaxy minor axes.

It is also possible that the observed radio emission 
does not accurately trace the outflow, but only represents regions of high
column density of synchrotron-emitting electrons, such as an edge
of an expanding bubble.  

It is, however, difficult to explain how a starburst-driven wind could produce
large-scale radio sources which are bi-symmetric with respect to the nucleus,
but are not oriented perpendicular to the major axis.  Only the scenario in
which the active nucleus is itself a powerful starburst (e.g. Terlevich 1992) 
and the `wind' is collimated by
a thick gas torus (such as is proposed in unified models for AGN,
cf. Antonucci 1993) would seem to apply.
Such bi-symmetric radio morphologies, like that in NGC~4235, 
are more suggestive of the presence of 
a collimated outflow from the active nucleus.

If star formation activity
produces a significant fraction of the radio power in 
Seyferts, one might expect the radio and far-IR luminosities to be correlated
as in starburst galaxies (e.g. Helou \etal 
1985).  For
the Seyferts in our complete sample (Table 1), $\mu$,
the logarithm of the ratio of 6~cm radio flux to 60$\mu$ flux, has mean
value and dispersion 2.3$\pm$0.4.  For our four starburst galaxies, the
correlation is much tighter (2.50$\pm$0.06), consistent with results for
a much larger sample of IR-bright galaxies from Condon \& Broderick (1988;
2.45$\pm$0.3).  As expected, $<\mu>$ is smaller for the Seyferts since
the AGN produces some of the radio emission.

By subtracting the power of the linear nuclear radio sources from the total
radio power,  Baum \etal (1993) calculated `extranuclear' radio powers and 
computed corrected values of $\mu$ (``$\mu_{XN}$'').
They found that mean value of $\mu$ then
shifted upward so that it was consistent with the mean value of
$\mu$ for starbursts
(although the Seyferts had larger dispersion in $\mu$ by a factor $\approx$ 3).
If we attempt a similar correction (substituting 
P$_{XN}$ for P$_T$), we find that the distribution of $\mu_{XN}$
also shifts upward (2.6$\pm$0.6; however, one must be cautioned that,
in our calculation we have also subtracted
any nuclear diffuse [D-type] emission that may have been present in the 
nuclear region).  Since the two distributions of $\mu$ and $\mu_{XN}$ for
our sample overlap considerably (and both include the 
starburst value $\mu =$ 2.5), we cannot suggest
(or deny) that P$_{XN}$ is produced by star formation activity.
A more conclusive test could be done by only subtracting radio powers
corresponding to nuclear radio sources produced by the active
nucleus (e.g. unresolved cores and sub-kpc linear structures).
Unfortunately, high-resolution
radio maps are not available for most of the Seyferts in our sample.

\section{Summary and Conclusions}

As part of our program to study large-scale outflows (LSOs) in Seyfert 
galaxies,
we have obtained deep radio continuum images at 4.9~GHz (6~cm) with the VLA
of ten galaxies from our complete sample of edge-on Seyfert galaxies 
(paper~I) and four edge-on starburst galaxies which are well-matched to
the Seyferts in radio luminosity, axial ratio and recessional velocity. 
These radio maps and previously published
maps from the literature are used to investigate the nature and origin
of kpc-scale radio structures in Seyferts.

We found that six of the ten
Seyferts observed have large-scale radio structure
extending $\gapprox$1 kpc from the nucleus.  
All four starburst galaxies have a very bright disk component and one
has evidence for a radio halo which extends out of the galaxy plane.  
In the Seyfert galaxies, most of the radio emission is, in general, 
concentrated in the nuclear region ($\lapprox$1 kpc).  In our starburst 
galaxies (which have comparable total radio luminosity), the surface brightness
of the outer disk is much brighter than that in the Seyferts.

We found luminous 
(6~cm radio powers $\sim$1$-$8 $\times$ 10$^{19}$ W~Hz$^{-1}$), unresolved
radio sources in the disks of five of the ten Seyfert galaxies.  Follow-up
studies are warranted to determine if these sources are radio supernovae or
groups of supernova remnants.

Large-scale 
radio sources in Seyferts typically extend out to radii $\sim$1$-$5 kpc and
have radio powers and radial extents similar to those of radio 
halos in starburst galaxies.  In general, the large-scale
radio sources are not oriented perpendicular to the major axis, but appear
to emerge from the galaxy disk at a skewed angle.
Their morphologies resemble diffuse emission from a collimated outflow 
originating in the nuclear region and not a 
spherical halo extending out along the galaxy minor axis.  The large-scale
radio sources are often (but not always) associated with optical emission-line
regions and kinematically disturbed gas.

Although the active nucleus could easily both produce the
relativistic plasma and provide sufficient energy for driving the outflow,
nuclear starbursts are also capable of doing so.  Our result that 
large-scale radio structures are not preferentially 
oriented perpendicular to the 
galaxy major axes suggests that LSOs are not spherical winds blowing
out of the
nuclear region.  The observed large-scale radio structures are most easily
understood in the context of a directed outflow (e.g., an AGN-driven jet)
that is somehow diverted away from the galaxy disk on scales $\gapprox$1 kpc.
Future observational studies of LSOs in individual Seyfert galaxies are needed
in order to distinguish between all of the scenarios that may be occurring
in the nuclear region.
Such studies may also shed light on the relationship (if any exists)
between classical starburst galaxies, powerful radio galaxies and
Seyfert galaxies.

\acknowledgments

E.J.M.C. would like to thank Michael Dahlem, Alan Roy, and Brian Rush
for providing helpful information and for useful discussions.
E.J.M.C. thanks the Director's Office of the Space Telescope Institute for
providing funding.  This research has made extensive use of the NASA/IPAC
Extragalactic Database (NED), which is operated by the Jet Propulsion
Laboratory, Caltech, under contract with NASA.  This paper represents a
portion of E.J.M.C.'s Ph. D. thesis, to be submitted in partial fulfillment
of the requirements of the Graduate School of the University of Maryland.

\vfill \eject
\centerline{\bf Appendix: The Seyfert Galaxy IRAS~13197$-$1627 is
MCG~$-$3-34-64}

\def\IRASX{IRAS~13197$-$1627}

In Figure A1, we show contour maps of the 4.9~GHz radio emission from the
two galaxies MCG~$-$3-34-63 and MCG~$-$3-34-64.
We find total radio fluxes of 93.4 mJy 
(5.0 $\times$ 10$^{22}$ W~Hz$^{-1}$ at an assumed distance of 67.2 Mpc)
and 1.3 mJy (7.1 $\times$ 10$^{20}$ W~Hz$^{-1}$) for
MCG~$-$3-34-63 and MCG~$-$3-34-64, respectively.
The central source in MCG~$-$3-34-63 has a flux of 
0.6 mJy (3.4 $\times$ 10$^{20}$ W~Hz$^{-1}$) and the source positioned
$\sim$15\asec\ (4.9 kpc) northeast of the nucleus has a flux of
0.15 mJy (8.1 $\times$ 10$^{19}$ W~Hz$^{-1}$).
The radio source in MCG~$-$3-34-64 is slightly resolved along P.A. 135\deg.

Our complete sample of edge-on Seyfert galaxies was constructed by 
cross-referencing various Seyfert catalogs with RC3 (de~Vaucouleurs \etal 1991;
see paper~I).  One of the
objects selected by this process was the warm FIR galaxy \IRASX, which is
listed in the Seyfert catalog of Huchra (1995).
This object was first identified as a Seyfert galaxy by de Grijp \etal (1985).
Several workers have associated this IRAS source with 
MCG~$-$3-34-63, an edge-on galaxy with axial ratio 3.2 (RC3).  We have 
since found that \IRASX\ is not associated with MCG~$-$3-34-63,
but is instead associated with the galaxy MCG~$-$3-34-64, 
positioned $\sim$1.8\amin\ southeast of MCG~$-$3-34-63.

We suspected this to be the case when we found a large disagreement between 
our observed 
radio fluxes for MCG~$-$3-34-63 and those listed in the literature from 
previous radio observations (which were
undoubtedly measurements of the stronger radio source MCG~$-$3-34-64).
The IRAS fluxes listed for \IRASX\ (which probably 
include emission from both galaxies) 
also implied excessively large values of $\mu$
(i.e. the FIR flux was comparatively very large).  We later learned
(M. de~Robertis 1995, private communication) that the nuclear spectrum
which identified the Seyfert galaxy \IRASX\ is a spectrum of 
MCG~$-$3-34-64, not MCG~$-$3-34-63.

Authors of future catalogs of Seyferts should note that this error has
progressed into many published catalogs and papers.

\clearpage
 
\begin{deluxetable}{rlcrrcrr}
\scriptsize
\tablewidth{7in}
\tablecaption{ {\bf Complete Statistical Sample of Edge-on Seyfert Galaxies}}
\tablehead{
\colhead{New} &
\colhead{Galaxy} & 
\colhead{Seyfert} &
\colhead{D\tablenotemark{a}} &
\colhead{P$_6$\tablenotemark{b}} & \colhead{Ref\tablenotemark{b}} &
\colhead{L$_{FIR}$\tablenotemark{c} } & \colhead{$\mu$\tablenotemark{d}} \\
\colhead{Map?} &
\colhead{Name} & 
\colhead{Type} &
\colhead{(Mpc)} &
\colhead{($\log$ W~Hz$^{-1}$)} & \colhead{} &
\colhead{($\log$ erg~s$^{-1}$)} & \colhead{} \\
} 
\startdata
      & IC~1657       & 2   & 47.4 & 21.35    &   1   & 43.69 & 2.52  \\
 \dag & UM~319        & 2   & 63.1 & 21.43    &   2   & 43.60 & 2.47  \\
 \dag & Mrk~993       & 2   & 62.1 & 21.03    &   2   & 43.08 & 2.10  \\
      & Mrk~577       & 2   & 69.1 & $<$19.84 &   2   & {...} & {...} \\
 \dag & Ark~79        & 2   & 68.8 & 21.55    &   2   & {...} & {...} \\
      & NGC~931       & 1   & 66.6 & 21.52    &   3   & 43.94 & 2.66  \\
 \dag & NGC~1320      & 2   & 36.0 & 20.74    &   2   & 43.22 & 2.80  \\
      & NGC~1386      & 2   & 20.0 & 21.01    &   3   & 43.18 & 2.45  \\
 \dag & NGC~2992      & 2   & 30.9 & 22.00    &   2   & 43.81 & 2.09  \\
 \dag & MCG~$-$2-27-9 & 2   & 62.0 & 20.65    &   2   & 43.21 & 2.69  \\
 \dag & NGC~4235      & 1   & 32.1 & 21.03    &   2   & 42.36 & 1.56  \\
      & NGC~4388      & 2   & 33.6 & 21.78    &   3   & 43.88 & 2.37  \\
 \dag & NGC~4602      & 1.9 & 34.0 & 20.92    &   2   & 43.65 & 2.90  \\
      & NGC~4945      & 2   &  6.7 & 22.35    &   4   & 44.02 & 1.94  \\
      & IC~4329A      & 1   & 63.9 & 22.22    &   3   & 43.68 & 1.81  \\
 \dag & NGC~5506      & 2   & 24.2 & 22.10    &   2   & 43.44 & 1.67  \\
      & ESO~103-G35   & 2   & 53.1 & {...}    & {...} & 43.48 & {...} \\
      & NGC~6810      & 2   & 26.1 & {...}    & {...} & 43.93 & {...} \\
 \dag & IC~1368       & 2   & 52.2 & 21.53    &   2   & 43.82 & 2.59  \\
      & IC~1417       & 2   & 57.5 & {...}    & {...} & 43.14 & {...} \\
      & NGC~7410      & 2   & 23.3 & 20.29    &   5   & 42.60 & 2.38  \\
      & NGC~7590      & 2   & 21.3 & 21.42    &   6   & 43.45 & 2.21  \\
\enddata

\tablenotetext{}{Optical positions, axial ratios and recessional velocities
for these objects are listed in paper~I, Table 1. }
\tablenotetext{\dag}{New radio maps are presented in this paper.}
\tablenotetext{a}{Assumed distance in Mpc.  Except for NGC~1386 and NGC~4945,
  distances were calculated from recessional velocities listed in RC3
  (see paper~I), using H$_o =$ 75.  The distance to NGC~1386 was taken as
  20.0 Mpc (the distance to the Fornax cluster) and a distance of 6.7 Mpc was
  assumed for NGC~4945.}
\tablenotetext{b}{Total radio power at 6~cm (4.89~GHz).  Fluxes measured at
   other frequencies were converted to fluxes at 4.89~GHz assuming
   S$_\nu \propto \nu^{-0.75}$.  References:
   (1) 1.42~GHz VLA B/C config, Unger \etal 1981;
   (2) this paper;
   (3) 4.89~GHz VLA D config, Rush \etal 1995;
   (4) 4.75~GHz Parkes, Harnett \etal 1989;
   (5) 1.49~GHz VLA D config, Condon 1987;
   (6) 2.70~GHz single dish, Wright 1974 }
\tablenotetext{c}{Far-infrared luminosities, determined from IRAS fluxes using
  the method described in Fullmer \& Lonsdale (1989).  Fluxes were taken from
  Rush, Malkan \& Spinoglio (1993) when available, otherwise from
  Moshir \etal 1990 (as listed in the NASA Extragalactic Database [NED]). }
\tablenotetext{d}{Logarithm of ratio of 60$\mu$ flux density to total 
  6~cm flux density.  }

\end{deluxetable}

\begin{deluxetable} {llrrrrrr}
\scriptsize
\tablecaption{ {\bf Comparison Sample of Edge-on Starburst Galaxies}}
\tablehead{
\colhead{Galaxy} & \colhead{Galaxy\tablenotemark{a}} &
\colhead{$\log R_{25}$\tablenotemark{b}} & \colhead{$cz$\tablenotemark{b}} &
\colhead{D\tablenotemark{b}} &
\colhead{P$_6$\tablenotemark{c}} &
\colhead{L$_{FIR}$\tablenotemark{d} } & \colhead{$\mu$\tablenotemark{e}} \\
\colhead{Name} & \colhead{Type} &
\colhead{} & \colhead{(km~s$^{-1}$)} &
\colhead{(Mpc)} &
\colhead{($\log$ W~Hz$^{-1}$)} &
\colhead{($\log$ erg~s$^{-1}$)} & \colhead{} \\
} 
\startdata
UGC~903  & S?            & 0.70 & 2518 & 33.6 & 21.42 & 43.75 & 2.57 \\
NGC~1134 & S?            & 0.46 & 3644 & 48.6 & 21.90 & 44.13 & 2.48 \\
NGC~3044 & SB(s)c? sp    & 0.84 & 1292 & 17.2 & 21.12 & 43.30 & 2.42 \\
NGC~7541 & SB(rs)bc: pec & 0.45 & 2681 & 35.7 & 21.95 & 44.24 & 2.52 \\
\enddata
\tablenotetext{a}{Morphological type, taken from the NASA 
  Extragalactic Database (NED).
}
\tablenotetext{b}{Axial ratios ($R_{25}$) were taken from RC3.
  Recessional velocities ($cz$) are optical (or 21~cm, when
  available) velocities listed in RC3, unless otherwise noted.
  Distances were calculated from recessional velocities, using H$_o =$ 75.  }
\tablenotetext{c}{Total radio power at 6~cm (4.89~GHz), measured from the
  radio maps presented in this paper.}
\tablenotetext{d}{Far-infrared luminosities, determined from IRAS fluxes using
  the method described in Fullmer \& Lonsdale (1989).  Fluxes were taken from
  Moshir \etal 1990 (as listed in the NASA Extragalactic Database [NED]). }
\tablenotetext{e}{Logarithm of ratio of 60$\mu$ flux density to total
  6~cm flux density. }
\end{deluxetable}

\begin{deluxetable} {lrrrcc}
\tablecaption{ {\bf Unresolved Radio Sources in the Disks of Seyfert Galaxies} }
\tablehead{
\colhead{Galaxy} & \colhead{R.A.} & \colhead{Dec.} & 
\colhead{F$_6$\tablenotemark{a}} & \colhead{P$_6$\tablenotemark{a}} \\
\colhead{Name}   & \multicolumn{2}{c}{(J2000)} & 
\colhead{($\mu$Jy)} & \colhead{(10$^{19}$ W~Hz$^{-1}$)} \\
} 
\startdata
UM~319        & 01 23 21.2 & $-$01 58 52 & 126 & 6.0 \\
Ark~79        & 02 17 24.1 & $+$38 24 53 & 135 & 7.7 \\
NGC~1320      & 03 24 48.0 & $-$03 02 19 &  95 & 1.5 \\
              & 03 24 49.5 & $-$03 02 46 &  67 & 1.0 \\
MCG~$-$2-27-9 & 10 35 28.0 & $-$14 07 58 & 105 & 4.8 \\
              & 10 35 29.5 & $-$14 07 47 &  83 & 3.8 \\
NGC~4602      & 12 40 33.6 & $-$05 07 46 &  99 & 1.4 \\ 
              & 12 40 34.2 & $-$05 07 34 & 331 & 4.6 \\ 
              & 12 40 34.3 & $-$05 07 45 &  99 & 1.4 \\ 
              & 12 40 35.4 & $-$05 07 47 & 276 & 3.8 \\ 
              & 12 40 35.9 & $-$05 07 38 & 292 & 4.0 \\ 
              & 12 40 36.0 & $-$05 07 56 & 240 & 3.3 \\ 
              & 12 40 36.2 & $-$05 07 39 & 326 & 4.5 \\ 
              & 12 40 36.8 & $-$05 07 41 & 235 & 3.3 \\ 
              & 12 40 37.2 & $-$05 07 44 & 126 & 1.7 \\ 
              & 12 40 37.9 & $-$05 08 07 & 144 & 2.0 \\ 
              & 12 40 38.1 & $-$05 07 45 & 457 & 6.3 \\ 
              & 12 40 38.8 & $-$05 07 60 & 149 & 2.1 \\ 
              & 12 40 39.2 & $-$05 08 12 & 571 & 7.9 \\ 
\enddata
\tablenotetext{a}{ Total flux and power at 6~cm, assuming the source is at the
  same distance as the galaxy.  Fluxes were measured from the normal maps
  by summing the flux density enclosed in boxes.}
\end{deluxetable}

\begin{deluxetable}{lrrrrcrrl}
\scriptsize
\tablewidth{7.25in}
\tablecaption{ {\bf Nuclear and Extranuclear Radio Components} }
\tablehead{
\colhead{Galaxy} & 
\colhead{F$_{T}$\tablenotemark{a}} & \colhead{F$_{N}$\tablenotemark{a}} &
\colhead{P$_{N}$\tablenotemark{b}} & 
\colhead{P$_{XN}$\tablenotemark{c}} & 
\colhead{P$_{LSO}$\tablenotemark{d}} &
\colhead{log R\tablenotemark{d}} & 
\colhead {$\mu_{XN}$\tablenotemark{e}} &
\colhead{Notes\tablenotemark{f}} \\
\colhead{Name} &
\colhead{(mJy)} & \colhead{(mJy)} &
\colhead{($\log$ W~Hz$^{-1}$)} &
\colhead{($\log$ W~Hz$^{-1}$)} & \colhead{($\log$ W~Hz$^{-1}$)} &
\colhead{} & \colhead{} &
\colhead{} \\
} 
\startdata
UM~319         &   5.7 &   4.0 & 21.28  & 20.92 & 20.9 & 0.4 & 3.00 &
  lin struct, src in disk \\
Mrk~993        &   2.3 &   1.9 & 20.95  & 20.30 & {...} & {...} & {...} &
  unr or v sl res \\
Ark~79         &   6.2 &   4.4 & 21.40  & 21.01 & 21.0 & 0.4 & {...} &
  sl res to SW, src in disk \\
NGC~1320       &   3.5 &   2.1 & 20.51  & 20.35 & 20.3 & 0.2 & 3.24 &
  lin struct, srcs in disk \\
NGC~2992       &  87.7 &  37.8 & 21.63  & 21.76 & 21.8 & -0.2 & 2.29 &
  lots of struct, extent min axis \\
MCG~$-$2-27-9  &   1.0 &   0.6 & 20.40  & 20.29 & {...} & {...} & {...} &
  extent to SE, srcs in disk \\
NGC~4235       &   8.6 &   4.6 & 20.75  & 20.70 & 20.7 & 0.1 & 1.89 &
  unr, v extended \\
NGC~4602       &   6.1 &   1.9 & 20.42  & 20.76 & {...} & {...} & {...} &
  weak nuc src, disk clump \\
NGC~5506       & 178.8 & 139.2 & 21.99  & 21.44 & 21.4 & 0.6 & 2.37 &
  lots of struct, extent min axis \\
IC~1368        &  10.4 &   8.9 & 21.46  & 20.70 & {...} & {...} & {...} &
  sl res along maj axis \\
\tableline 
UGC~903        &  19.7 &   4.0 & 20.73  & 21.32 & {...} &  {...} & {...} &
  disk clump, fingers out of disk \\
NGC~1134       &  27.9 &   4.0 & 21.05  & 21.83 & {...} &  {...} & {...} &
  disk clump \\
NGC~3044       &  37.0 &   1.9 & 19.84  & 21.10 & 19.7  & {...} & {...} &
  disk clump, halo \\
NGC~7541       &  58.5 &  10.5 & 21.21  & 21.87 & {...} & {...} & {...} &
  disk clump, fingers \\
\enddata
\tablenotetext{}{The ten edge-on Seyfert galaxies are listed first and the 
  four starburst galaxies are listed next.}
\tablenotetext{a}{6~cm fluxes.  Total flux from the galaxy F$_{T}$ was 
  measured by summing the flux density inside a box enclosing all of the 
  emission from the galaxy.  Nuclear flux F$_{N}$ was estimated from the
  peak flux density in the nuclear region, using the ``hi-res''
  maps (see section 2).}
\tablenotetext{b}{Nuclear 6~cm radio powers, calculated from F$_{N}$.  }
\tablenotetext{c}{Extranuclear (XN) 6~cm radio powers.  The extranuclear
  power P$_{XN}$ was 
  calculated directly from F$_{XN} =$ F$_{T} -$ F$_{N}$.  }

\tablenotetext{d}{ The extranuclear 6~cm radio powers associated with the 
  large-scale outflows (LSOs).  Except for NGC~3044, P$_{LSO}$ was calculated 
  by subtracting contributions from sources in the disk from P$_{XN}$.
  For NGC~3044, flux of the radio halo (roughly estimated from Figure 2c)
  was used to calculate P$_{LSO}$. }
\tablenotetext{e}{Logarithm of ratio of 60$\mu$ flux density to extranuclear
  6~cm flux density F$_{XN}$. }
\tablenotetext{f}{Notes about radio structure.  For the Seyfert galaxies, 
  the nuclear structure is first described, then the extranuclear structure.
  Abbreviations: lin = linear, struct = structure, src = source,
  min = minor, unr = unresolved, v = very, nuc = nuclear
  sl = slight(ly), res = resolved, maj = major
}
\end{deluxetable}

\begin{deluxetable}{lrrrcrr}
\scriptsize
\tablecaption{{\bf Geometrical Properties of Large-Scale Radio Emission in 
Seyfert Galaxies }}
\tablehead{
\colhead{Galaxy} & 
\colhead{P.A.(LSO)\tablenotemark{a}} &
\multicolumn{2}{c}{Max Radius\tablenotemark{b}} &
\colhead{Ref\tablenotemark{c}} &
\colhead{P.A.(MAJ)\tablenotemark{d}} &
\colhead{$\Delta$\tablenotemark{e}} \\
\colhead{Name} &
\colhead{(deg)} &
\colhead{(arcsec)} & \colhead{(kpc)} &
\colhead{} &
\colhead{(deg)} &
\colhead{(deg)} \\
} 
\startdata
UM~319        & 130 &  10 &  3.2 & 1 & 165 & 35 \\         
Ark~79        & 200 &   6 &  1.9 & 2 &  87 & 67 \\
NGC~1320      & 180 &   6 &  1.1 & 2 & 135 & 45 \\
NGC~2992      & 105 &  25 &  3.7 & 3 &  15 & 90 \\
NGC~4235      & 275 &  62 &  9.7 & 4 &  48 & 47 \\
NGC~4388      &   0 &  22 &  3.6 & 3 &  92 & 88 \\
NGC~4945      & 140 & 960 & 31.4 & 5 &  43 & 83 \\
IC~4329A      & 285 &   6 &  1.9 & 6 &  45 & 60 \\
NGC~5506      & 145 &  38 &  4.4 & 4 &  91 & 54 \\
\enddata
\tablenotetext{a}{Position angle of large-scale radio structure}
\tablenotetext{b}{Maximum projected radius of large-scale radio emission along 
   P.A. listed in column 2.}
\tablenotetext{c}{References for maps used to measure P.A. (column 2) 
and projected radius (columns 3 and 4):
(1) this paper, normal map;
(2) this paper, hi-res map;
(3) Hummel \etal (1983) -- see Figures 1e and 1h;
(4) this paper, tapered map;
(5) Harnett \etal (1989) -- see Figure 1j;
(6) Unger \etal (1987) -- see Figure 1k }
\tablenotetext{d}{Position angle of the major axis of the galaxy, taken from
  RC3 or measured from the digitized optical sky-survey plates.}
\tablenotetext{e}{Difference in angle between major axis and 
  P.A. of large-scale radio emission}
\end{deluxetable}

\begin{deluxetable}{lcc}
\tablecaption{ {\bf Evidence for Large-Scale Outflows in Edge-on 
Seyfert Galaxies}}
\tablehead{
\colhead{Galaxy} & 
\colhead{Optical} &
\colhead{Radio} \\
\colhead{Name} &
\colhead{Evidence\tablenotemark{a}} &
\colhead{Evidence\tablenotemark{b}} \\
} 
\startdata
UM~319           & {...} &   XN \\
Ark~79           &   S?  &   XN   \\
NGC~1320         &  none &   XN   \\
NGC~2992         &   I   &   XP  \\
NGC~4235         &  none &  XP  \\
NGC~4388         & I,S   &  XN  \\
NGC~4945         & I,S   &   XP \\
IC~4329A         &  I    &   XN \\
NGC~5506         &  I,S  &   XP  \\
IC~1368          &  I    & none \\
\enddata
\tablenotetext{a}{Evidence for minor-axis outflows from paper~I.  Blank
  entry denotes insufficient data, whereas ``none'' denotes no evidence.
  `I' denotes evidence from \han2 images and
  `S' denotes evidence from long-slit spectra.  Question mark indicates
  only suggestive evidence.  }
\tablenotetext{b}{Evidence for large-scale outflows from the present radio 
  maps.  `XN' denotes large-scale ($\gapprox$ 1 kpc) extranuclear radio
  emission, `XP' denotes that emission clearly
  extends out of the optical disk of the galaxy (see Figures 1), and
  ``none'' denotes no evidence.  }
\end{deluxetable}

\clearpage

\noindent Fig. 1.-- 
{\bf Large-scale radio maps of edge-on Seyfert galaxies:}
Contour maps of the radio emission overlaid on on grayscale plots of 
the optical emission.  The optical images are digitized versions
of optical
survey plates from the STScI digitized sky survey.  
See section 2 for a description of the `normal' 
(beam size 5\asec\ $\times$ 5\asec\ FWHM), `hi-res' 
(3\asec\ $\times$ 3\asec\ FWHM) and `tapered' 
(10\asec\ $\times$ 10\asec\ FWHM) 6~cm maps.
Unless otherwise stated, contour levels are 25 $\mu$Jy $\times$
-3, 3, 5, 7 and 10 and then successive levels are a factor of 2 larger.
{\bf (a) UM~319. }  Normal (left) and hi-res (right) maps.  
Peak flux densities are 4.4 and 4.0 mJy~beam$^{-1}$, respectively.
{\bf (b) Mrk~993. } Normal map.
The peak flux density is 2.9 mJy~beam$^{-1}$.
{\bf (c) Ark~79. } Normal (top) and hi-res (bottom) maps.
Peak flux densities are 5.2 and 4.4 mJy~beam$^{-1}$, respectively.
{\bf (d) NGC~1320. } Normal (left) and hi-res (right) maps.
Peak flux densities are 2.4 and 2.0 mJy~beam$^{-1}$, respectively.
{\bf (e) NGC~2992. } Normal (left) and tapered (middle) maps plus reproduction
of contours from the 20~cm map (6\asec\ beam)
from Hummel \etal (1983, right).  Peak flux densities in the normal and 
tapered maps
are 49.3 and 68.4 mJy~beam$^{-1}$, respectively.   Contour levels in the 
20~cm map are -2, 2, 4, 7, 10, 20, 30, 50, 70, 90, 110 mJy~beam$^{-1}$
{\bf (f) MCG~$-$2-27-9. } Normal map.
The peak flux density is 4.5 mJy~beam$^{-1}$.
{\bf (g) NGC~4235. } Normal (top) and tapered (middle) maps plus 
20~cm map (beam 58.7\asec\ $\times$ 51.1\asec\ FWHM, P.A. $+$72\deg)
from Edelson (1987) data (lower). 
Peak flux densities in the normal and tapered maps
are 5.3 and 5.6 mJy~beam$^{-1}$, respectively.   Contour levels in the 
20~cm map are 0.4 mJy~beam$^{-1}$ $\times$ ( 3, 5, 7, 10 and 20).
{\bf (h) NGC~4388. } Reproduction of the contours from the 6~cm map 
(beam 5.6\asec\ $\times$ 5.1\asec, P.A. $-$20\deg)
from Hummel \etal (1983).
Contour levels are 
0.25, 0.5, 1, 2, 4, 
8, 
12 
and 16 mJy~beam$^{-1}$.
{\bf (i) NGC~4602. } Normal map.  
The peak flux density is 2.0 mJy~beam$^{-1}$.
{\bf (j) NGC~4945. } Reproduction of the contours from the 4.75~GHz 
(6~cm; 4.6\amin\ beam)
map from Harnett \etal (1989).
Contour levels are 
0, 10, 25, 40, 
70, 
100, 250, 500, 750, 1000, 1250, 1500, 1750, and 2000 mJy~beam$^{-1}$.
{\bf (k) IC~4329A. } Reproduction of contours from the 6~cm radio map 
(beam size 1.3\asec\ $\times$ 1.2\asec, P.A. $-$31\deg)
of the radio emission from Unger \etal (1987).
Contour levels are 19 mJy~beam$^{-1}$ $\times$ percentages 
(2, 4, 6, 8, 16, 30, 40, and 50).
{\bf (l) NGC~5506. } Normal (top) and tapered (bottom) maps.
Peak flux densities are 159.9 and 170.4 mJy~beam$^{-1}$, respectively.
{\bf (m) IC~1368. } Normal map.
The peak flux density is 9.8 mJy~beam$^{-1}$.
{\bf (n) NGC~7410. } 20~cm map (48\asec\ beam) from Condon (1987).
Contour levels are 0.1 mJy~beam$^{-1}$ $\times$ ( -3, 3, 5, 7, 10, 20, 40).
The peak flux density is 5.4 mJy~beam$^{-1}$.

\noindent Fig. 2.--
{\bf Large-scale radio maps of edge-on starburst galaxies:}
As in Figure 1, 
contour maps are overlaid on on grayscale plots of 
the optical emission (images from the
STScI digitized sky survey).  
See section 2 for a description of the `normal' and `tapered' 6~cm maps.
Unless otherwise stated, contour levels are 25 $\mu$Jy $\times$
-3, 3, 5, 7 and 10 and then successive levels are a factor of 2 larger.
{\bf (a) UGC~903. } Normal map.
The peak flux density is 5.0 mJy~beam$^{-1}$.
{\bf (b) NGC~1134. } Normal map.
The peak flux density is 5.4 mJy~beam$^{-1}$.
{\bf (c) NGC~3044. } Normal (top) and tapered (middle) maps plus reproduction
of the 20~cm map (18\asec\ beam) from Condon \etal (1990, bottom).
Peak flux densities in the normal and tapered maps
are 3.2 and 6.6 mJy~beam$^{-1}$, respectively.   The lowest contour levels in 
the 20~cm map is 0.5 mJy~beam$^{-1}$ and successive levels are a factor of
$\sqrt{2}$ larger in flux density.
{\bf (d) NGC~7541. } Normal map.
The peak flux density is 12.1 mJy~beam$^{-1}$.

\noindent Fig. 3. --
Histogram of the maximum projected radius of large-scale radio structures 
in Seyferts.  Values were taken for the nine Seyferts listed in Table 4.

\noindent Fig. 4. --
Orientation of the large-scale radio structures in Seyferts.
Histogram of $\Delta$, the angle between the large-scale emission and
and the galaxy major axis.  
Values were taken for the nine Seyferts listed in Table 4.

\noindent Fig. A1 -- Normal (see section 2) radio map overlaid on a grayscale
plot of an optical image from the STScI digitized sky survey.  The edge-on
galaxy MCG~$-$3-34-63 is located in the upper right of the image, whereas
the Seyfert galaxy MCG~$-$3-34-64 is located in the lower left.  Contour
levels are 25$\mu$Jy $\times$ 
-3, 3, 5, 7 and 10 and then successive levels are a factor of 2 larger.
The peak flux density is 89.9 mJy~beam$^{-1}$.

\end{document}